\documentclass[aps,twocolumn,floats,prx]{revtex4}
\usepackage{graphics,graphicx,epsfig}
\usepackage{amssymb,color}
\usepackage{amsmath}
\usepackage{epsf,epstopdf,wrapfig}
\usepackage{amsmath,textgreek,bm}
\usepackage{hyperref}

\usepackage[utf8]{inputenc}
\usepackage[T1]{fontenc}

\usepackage{marginnote}
\usepackage{mathtools}

\begin{document}

\title{Large language models and the entropy of English}

\author{Colin Scheibner$^{a,b,\dagger}$}
\author{Lindsay M. Smith$^{a,d,\dagger}$}
\author{William Bialek$^{a,c,d}$} 
\affiliation{$^a$Joseph Henry Laboratories of Physics, $^b$Princeton Center for Theoretical Science, and $^c$Lewis--Sigler Institute for Integrative Genomics, Princeton University, Princeton NJ 08544 USA\\
$^d$Initiative for the Theoretical Sciences, The CUNY Graduate Center, 365 Fifth Avenue, New York NY 10016 USA} 

\begin{abstract}
We use large language models (LLMs) to uncover long--ranged structure in English texts from a variety of sources. The conditional entropy or code length in many cases continues to decrease with context length at least to $N\sim 10^4$ characters, implying that there are direct dependencies or interactions across these distances. A corollary is that there are small but significant correlations between characters at these separations, as we show from the data independent of models.  The distribution of code lengths reveals an emergent certainty about an increasing fraction of characters at large $N$. Over the course of model training, we observe  different dynamics at long and short context lengths, suggesting that long--ranged structure is  learned only gradually.  Our results constrain efforts to build statistical physics models of LLMs or language itself.
\end{abstract}

\date{\today}
\maketitle

\begingroup
\renewcommand\thefootnote{$\dagger$}
\footnotetext{These authors contributed equally to this work.}
\endgroup

Entropy helps us to describe phenomena ranging from steam engines to black holes.  Entropy also quantifies our intuitive notion of information, and this measure is unique in satisfying plausible constraints \cite{Shannon_1948}.  Shannon used language as an accessible example of these ideas, and played a ``guessing game'' with human subjects to estimate the entropy of written English \cite{Shannon_1951}.  The results of this experiment gave bounds on the conditional entropy for a single character in the text given knowledge of the previous $N$ characters; tighter bounds can be obtained by asking subjects not just to guess but to bet \cite{Cover+King_1978}.  Decades later, performance at the guessing game is essentially the objective function for training large language models.

Shannon interpreted his results as suggesting that the conditional entropy per character approached a plateau for $N\sim 100$.  Hilberg argued that the data are actually consistent with a decay $\sim 1/\sqrt{N}$ \cite{Hilberg_1990}; if this continues to larger $N$, then the entropy of texts would be sub--extensive \cite{Ebeling+Nicolis_1992,Debowksi_2011}.  There are hints that the mutual information between characters decays as a power of their separation  and that the number of distinct words in long texts grows as a fractional power of their length \cite{Ebeling+Poschel_1994}.  These could all be signs of long range, scale-invariant correlations, evoking connections to statistical physics.   
But these data were limited to modest $N$, and  there is a long history of skepticism about whether statistical structure tells us anything interesting about language \cite{Chomsky_1956}.

Large language models (LLMs) give us a new tool to explore the entropy of text.  In particular, we can play Shannon's  guessing game, asking the model to generate the next character (or, more precisely, the next token) given a sequence of $K$ tokens from real text.  Since the model returns the full distribution rather than a single guess, we can compute directly the model's estimate of the conditional entropy.  We also can use the model to encode the actual next token, and this code length bounds the conditional entropy of the real text.  Importantly we can do these computations out to $K\sim 10^3 - 10^4$.  Along the path to  these models the community has assembled enormous data sets from which we can estimate the decay of correlations out to similarly large separations.  

Here we show that across many classes of text the code lengths generated by well-trained LLMs agree with one another and continue to decrease at scales $N>  10^3$, in some cases with no sign of a plateau. This is possible only if there are effective interactions that reach across these long distances.  A corollary is that there is a small but significant mutual information between characters at these large separations.  Going beyond the mean,  we see structure in the distribution of code lengths including an approximately power--law divergence in the distribution at small code lengths and large $N$, pointing toward an emergence of near certainty about the next character in a long sequence. Finally we explore how these large $N$ behaviors develop as the models learn.  We present a brief overview of these results  and discuss their implications.  A fuller account will be given elsewhere \cite{SSB2}.

LLMs are trained on ensembles of text obtained by scraping the internet. These corpora can be used for the study of the structure of language itself. 
We use a publicly available example, the English variant of the Colossal Clean Crawled Corpus (C4), a collection curated from $> 3.65 \times 10^8$ internet documents using automated filters~\cite{raffel2019exploring,dodge2021documenting}. 
More focused data sets include a collection of pages from English Wikipedia and Simple English Wikipedia~\cite{wikidump}, as well as news articles from the BBC~\cite{li2024latesteval}; the last are especially useful since they are time stamped, so we can test models with text that could not have contributed to their training. We also study a corpus of narratives paired with their summaries and analyses~\cite{kryscinski2021booksum}.
Finally, we analyze poetry drawn from two sources: the Gutenberg Poetry Corpus, a collection of poetry mined from the Project Gutenberg book collection~\cite{parrish_gutenberg_poetry_corpus}, and a more carefully curated collection from the Poetry Foundation~\cite{suayptalha_poetry_foundation_poems_hf}. For details see Appendix \ref{app:corpora}.

We use four models for our analysis: OLMo~2~1B~\cite{olmo20242olmo2furious}, Llama~3.2~1B~\cite{grattafiori2024llama3herdmodels}, Qwen3~8B~\cite{qwen3technicalreport}, and a 1.7B parameter model we train on a subset of the DCLM dataset~\cite{li2024datacomplm}. Each model uses a different tokenizer to convert text at the character level to discrete tokens. All models are open-source in their model weights (parameters);  OLMo~2 and our 1.7B  model have the advantage of being trained with open datasets as well. All models consist of a decoder-only Transformer \cite{vaswani2017attention}. We note that we trained DCLM~1.7B  on at least 2 orders of magnitude less data than the models released by large labs, and the maximum context length that it accommodates is 2048 tokens; this model also has no mid- or post-training. Despite this, we still see similar behaviors in the code length across context length. For details see Appendix \ref{app:models}.

Consider a segment of text, defined as a sequence of characters $\{c_1,\, c_2,\, \cdots ,\, c_N\}$.  LLMs start by converting this into a sequence of tokens $\{t_1,\, t_2,\, \cdots ,\, t_K\}\equiv t_{1\cdots K}$, where the average ratio of characters per token $N/K\sim 4-5$ is specific to each model and text corpus (Table \ref{tab:KtoN}, Appendix \ref{app:data}).  Given this input LLMs return the probability distribution of the next token, $P_K(t_{K+1}|t_{1\cdots K})$; we recall that this distribution provides a basis for encoding the next token with a code length \cite{Shannon_1948,Mezard+Montanari_2009}
\begin{equation}
\ell(t_{1\cdots K}) = -\log P_K(t_{K+1}|t_{1\cdots K}).
\end{equation}   
The mean code length
\begin{equation}
L (K) = \langle \ell (t_{1\cdots K})\rangle_{\rm data} = - \langle \log P_K(t_{K+1}|t_{1\cdots K})\rangle_{\rm data} 
\label{Ldef}
\end{equation}
is an upper bound on the conditional entropy of the real distribution out of which the text is drawn. The sum over sequence length bounds the total entropy
\begin{equation}
S(K) \leq \sum_{k=1}^K L(k), \label{eq:loss}
\end{equation}
and this is proportional to the loss function typically used for training the model, e.g.~with $K = 4096$ for the OLMo~2~1B pre-training.

\begin{figure}[t]
\centerline{\includegraphics[width = 0.9 \linewidth]{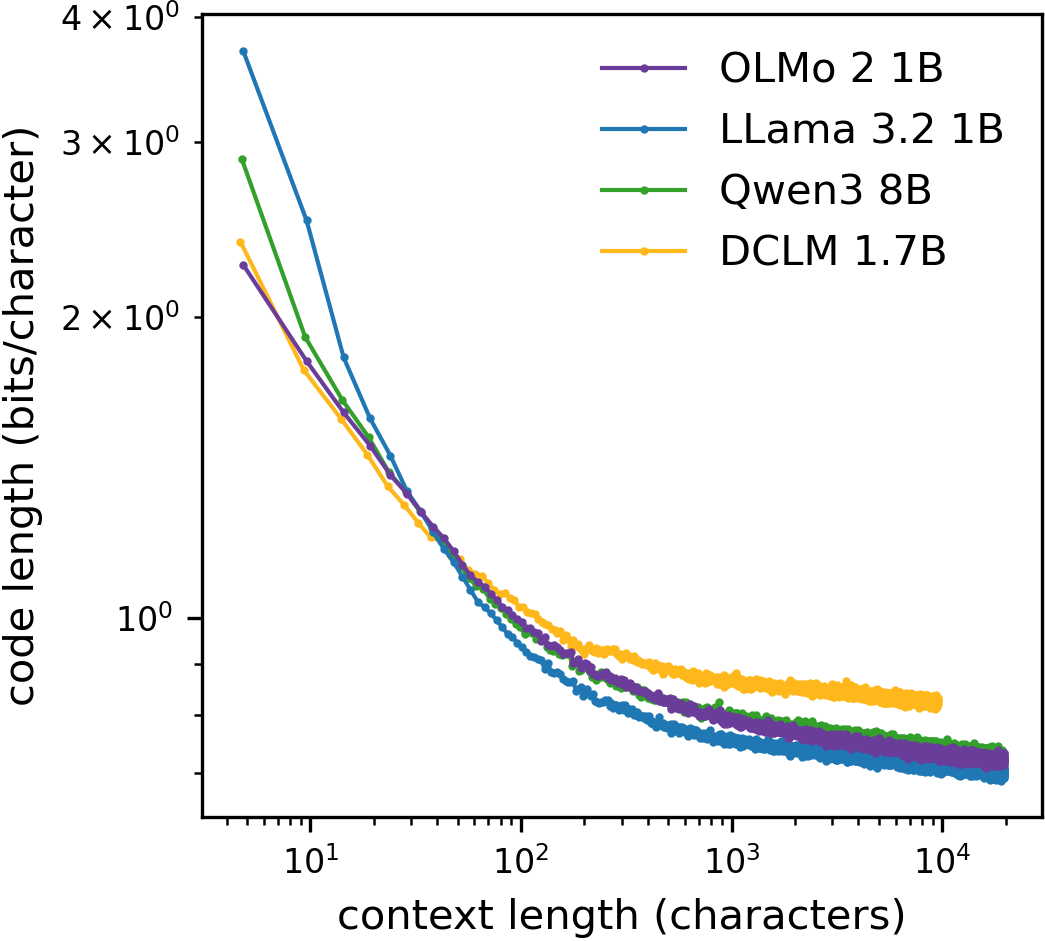}}
\caption{{\bf Code length vs. context length across models.} We evaluate $L(N)$ from Eq (\ref{Ldef}) for the C4 corpus using the OLMo~2 1B \cite{olmo20242olmo2furious}, Llama 3.2 1B \cite{grattafiori2024llama3herdmodels}, Qwen3 8B \cite{qwen3technicalreport}, and DCLM~1.7B~\cite{li2024datacomplm} models.  While there are differences of detail, all of these well trained models yield remarkably similar results. Error bars computed from the variance across random subsets of the data are smaller than the ``hash'' from point--to--point variability. \label{fig01}}
\end{figure}

Figure \ref{fig01} shows the code length $L(N)$ for the C4 corpus in the OLMo~2 1B~\cite{olmo20242olmo2furious}, Llama~3.2~1B~\cite{grattafiori2024llama3herdmodels}, Qwen3~8B~\cite{qwen3technicalreport}, and DCLM~1.7B~\cite{li2024datacomplm} models.  To make comparisons across models meaningful, we use characters rather than tokens as the unit of length, converting via the mean $N/K$ for each model.  We note that these results are within Shannon's bounds at $N=100$ \cite{Shannon_1951}, but the code length continues to fall slowly out to the largest $N\sim 10^4$.   

Three of the four models agree almost perfectly for $L(N > 10^3)$, and none show signs of a plateau at large $N$.  This is consistent with the conjecture that $L(N \rightarrow \infty)$ might actually vanish \cite{Hilberg_1990}, though the decay is much slower than one would estimate from data at smaller $N$.  It is perhaps surprising that the models disagree so much for $N < 100$. The objective function used in training is the total code length for strings of thousands of tokens, which certainly emphasizes large $N$.  But the models apparently can succeed in compressing long texts even while missing some of the small $N$ structure that we expect arises from rules of spelling and grammar.  Training of the smaller DCLM~1.7B model was entirely in our hands, and performance is not as good as for the models trained by larger groups, but the qualitative behavior is very similar.  The continuing decay at large $N$ is a bit slower and we cannot follow it quite as far because the model doesn't accommodate the longer contexts.

We can also compute the conditional entropy  
\begin{equation}
{\cal S}_{\rm cond}(t_{1\cdots K}) = - \sum_t P_K(t |t_{1\cdots K})\log_2 P_K(t|t_{1\cdots K}),
\label{Scond1}
\end{equation}
which measures the model's uncertainty in next-token prediction independent of the actual next token.  For three of the four models the mean conditional entropy  
\begin{equation}
s(K) = \langle {\cal S}_{\rm cond}(t_{1\cdots K}) )\rangle_{\rm data} 
\end{equation}
and the mean code length $L(K)$ agree closely at large $K$; see Fig.~\ref{fig:fig2Tokens} in Appendix \ref{app:data}.  For Qwen3, the conditional entropy is systematically smaller than the  code length, yet the code lengths remain in good agreement with those of the other models.   Agreement between the average entropy and code length would happen if the models were very good approximations to the true distribution, but this is not the only explanation.

\begin{figure}[t]
\centerline{\includegraphics[width =0.9\linewidth]{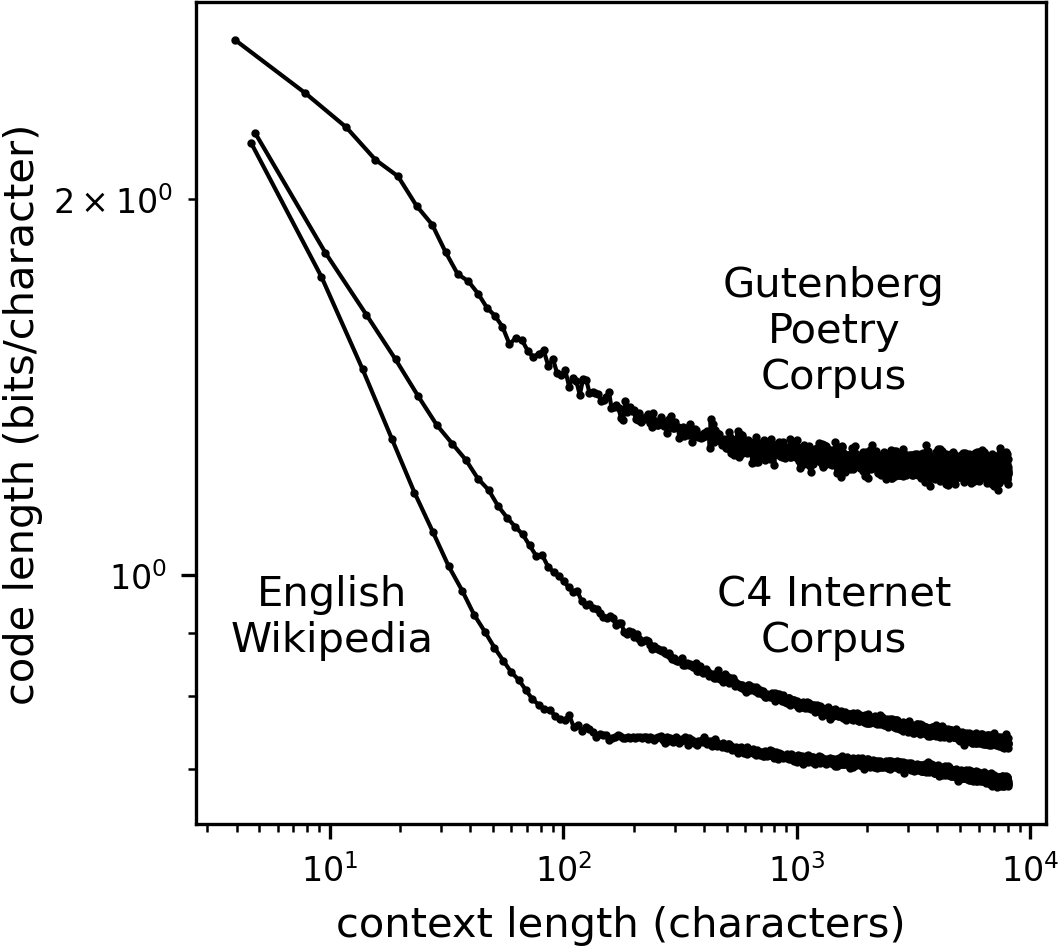}}
\caption{ {\bf Code length across genres.}~We evaluate $L(N)$ from Eq.~(\ref{Ldef}) via the OLMo~2~1B model over three text corpora: the C4 internet corpus (as in Fig.~\ref{fig01}), English Wikipedia~\cite{wikidump}, and the Gutenberg Poetry Corpus~\cite{parrish_gutenberg_poetry_corpus}.}
\label{fig02}
\end{figure}

Figure \ref{fig02} compares code lengths across multiple corpora, in each case using the OLMo 2 1B model; results with other models are similar as expected from Fig.~\ref{fig01} (see also Appendix~\ref{app:data}).  We see that code lengths are significantly longer in poems than in either the C4 corpus or the English Wikipedia pages.  The poems also seem to  approach a genuine plateau, while the other texts clearly have the code length decreasing slowly but significantly at $N\gg 10^3$.

\begin{figure*}[t]
\centerline{\includegraphics[width = \textwidth]{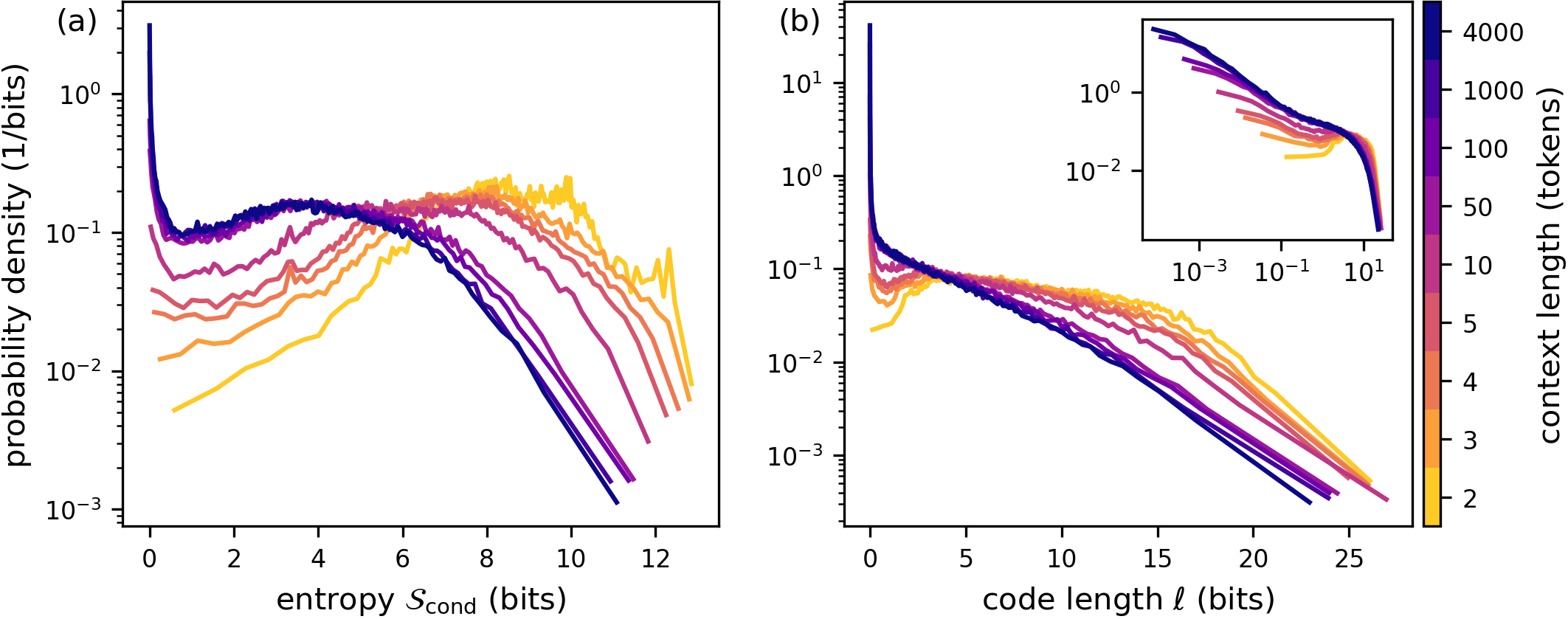}}
\caption{ {\bf Distribution of conditional entropy and code length.} The distribution of conditional entropy (a) and code length (b) across text samples evolves with the context length $K$.  Data from the C4 corpus as seen through the OLMo~2 1B model as in Fig.~\ref{fig02}. Results here are per token rather than per character; the OLMo~2~1B tokenizer uses 100,278 distinct tokens, with a mean number of characters per token $N/K = 4.79$ on the C4 dataset. Inset in (b) shows a log--log plot of the distribution, highlighting the near power--law tail at $\ell \rightarrow 0$. 
\label{fig03}}
\end{figure*}

The continuing decay of $s(K)$ and $L(K)$ at large $K$ is seen in the average over sample text strings, but the distributions of these quantities also change with $K$ (Fig.~\ref{fig03}).  As $K$ increases, a peak near zero entropy emerges from the bulk, indicating a kind of ``emergent certainty'' where the next token becomes---in some fraction of cases---almost perfectly predictable as we see more and more of the text.  In contrast, we see no sign that local rules of spelling or grammar induce this certainty at smaller $K$.  In addition to the growing peak near ${\cal S}_{\rm cond} = 0$, the bulk of the distribution shifts slowly to smaller values at increasing $K$, and these two effects make roughly equal contributions to the decline in average entropy.  Similar results are obtained across all corpora and models. Near certainty in next-token prediction has also been observed in synthetic text generated by LLMs~\cite{wang2025beyond}.

While the mean entropy and code length are quite similar for all genres across a large range of $K$, the distributions are quite different.  We see that the peak in the bulk distribution of ${\cal S}_{\rm cond}$ is replaced by a nearly exponential tail toward large $\ell$.  The behavior at small $\ell$ approximates a shallow power--law divergence of the distribution, and this becomes more accurate at larger context lengths (Fig.~\ref{fig03}b).

\begin{figure}[b]
    \centering
    \includegraphics[width=0.9\linewidth]{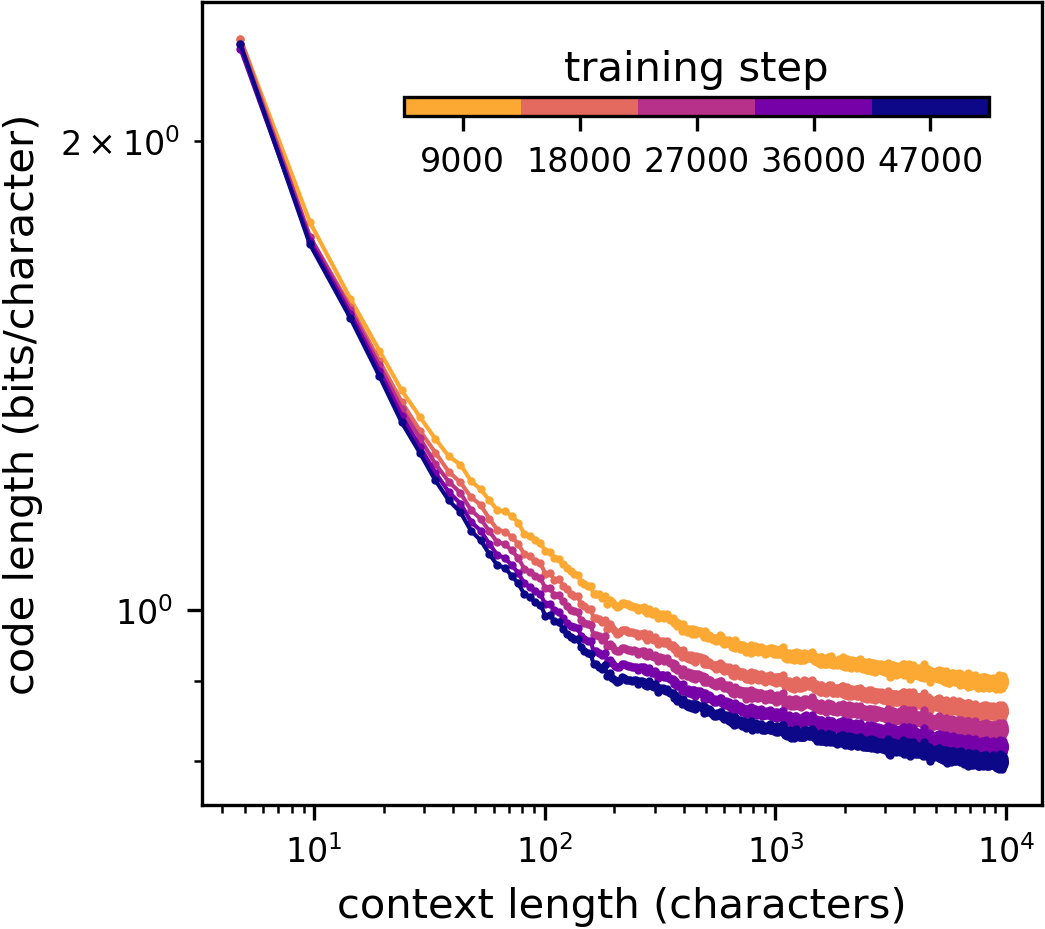}
    \caption{{\bf Development of code lengths during learning.}  Results for the DCLM 1.7B model \cite{li2024datacomplm}, at approximately equal intervals of training toward the final model.  Note the greater logarithmic decrease in code lengths at large context length.   
    Error bars are smaller than the hash, as in Fig \ref{fig01}.  
    \label{fig04}}
\end{figure}

The behaviors we see at large $N$ develops as models are trained.  We explore this in the DCLM 1.7B model \cite{li2024datacomplm} where we have complete control over the training. Figure \ref{fig04} shows that the models quickly reach nearly asymptotic values for $L(N)$ at small $N$ but the performance at large $N$ improves much more slowly.  This is a small but significant effect, and means that the model slowly learns to exploit longer and longer range dependencies.

A corollary of the decreasing code length at large $N$ is the presence of long--ranged correlations in the text. For large enough corpora, we can search for these correlations explicitly by estimating the mutual information between characters separated by a distance $d$ in the text.  Concretely, we look at  continuous strings of  characters chosen from a single corpus and accumulate the joint distributions  
\begin{equation}
Q_d(c,c') = \langle \delta_{c,c_n}\delta_{c',c_{n+d}}\rangle_{\rm data} .
\end{equation}
The mutual information then is 
\begin{equation}
I(d) = \sum_{c,c'} Q_d(c,c') \log_2\left[ {{Q_d(c,c')}\over{Q(c)Q(c')}}\right],
\label{Id_def}
\end{equation}
where $Q(c)$ is the (marginal) probability of seeing the individual character $c$.  The alphabet includes $n_c = 93$ characters---upper and lower case letters, numbers, punctuation, common symbols, and spaces (Appendix \ref{app:corpora}).  Information estimates are notoriously data hungry, and we expect systematic errors $\sim n_c^2/n_s$, where $n_s$ is the number of samples \cite{Bialek_2012}; in practice we have $n_s \sim 10^9$, so the floor of our measurements is $I\ll 10^{-4}\,{\rm bits}$. 

\begin{figure}[b]
\centerline{\includegraphics[width =0.9 \linewidth]{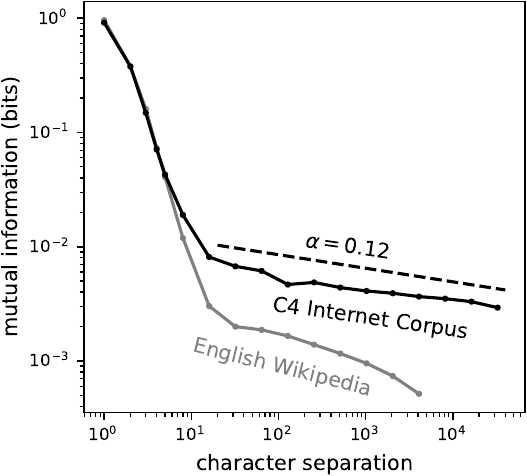}}
\caption{ {\bf Mutual information between characters as a function of separation.}  We show $I(d)$ from Eq.~(\ref{Id_def}) for the C4 and English Wikipedia corpora.  Errors computed from the variance across fractions of the data are smaller than the symbols. }
\label{Id}
\end{figure}

Figure \ref{Id} shows $I(d)$ for the two cases where we have enough data to avoid any sampling problems, C4 and English Wikipedia.  There are strong short--ranged correlations, decaying over   $\xi \sim 5-10$ characters, but there are residual long--ranged correlations that extend out to thousands of characters.  For the C4 corpus the behavior at large $d$ is consistent with a power--law $I(d) \propto d^{-\alpha}$, with $\alpha \approx 0.12$, and this is valid over at least two decades.  For the English Wikipedia data, the mutual information is smaller and there are signs of a more abrupt decay beyond $N\sim 10^3$.  These results are consistent with those in Fig \ref{fig02}, where we see that decay of $L(N)$ at large $N$ is faster in the C4 data.

It is a familiar result from statistical mechanics that correlations can extend over longer distances than interactions.  Thus the long--ranged correlations  measured by $I(d)$ do not imply direct dependence or interaction between characters separated by $d$, although the approximate power--law behavior for the C4 corpus is suggestive, since in simple one--dimensional models power--law correlations require power--law interactions \cite{Lieb+Mattis_1966}.  

The joint probability distribution for all the characters $\bm{c}\equiv \{c_n\}$ in a long string can be written as
\begin{equation}
    P(\bm{c}) = {1\over Z}\exp\left[-\sum_{n=1}^{R} V_n(\bm{c})\right],
\end{equation}
where each $V_n(\bm{c})$ involves interactions (nonlinear terms) among characters separated by at most $n$.  If the maximal range $R$ is finite, then $L(N>R)$ will be independent of $N$.  The continuing decrease of $L(N)$ seen in Figs.~\ref{fig01} and \ref{fig02} thus provides evidence for direct interactions across separations of nearly $10^4$ characters, with no obvious cutoff.  If the decay of $L(N)$ continues to arbitrarily large $N$ then the distribution $P(\bm{c})$ would have a sub--extensive entropy or vanishing entropy per character.

An alternative is a mean--field like model in which local groups of tokens interact with an auxiliary field or latent variable that is common across the whole length of the string.  The mean--field picture generalizes the classical ``bag of words'' model \cite{Harris_1954} to an ensemble of bags parameterized by continuous variables; the code length $L(N)$ declines with $N$ because we can estimate the underlying variable (i.e. identify the bag) with increasing precision as we see more text. But the absence of a plateau, at least for some corpora, means that the entropy of words drawn from a single bag, at fixed auxiliary field, must be very small, less than half a bit per character or just a handful of bits per word. Intermediate between restricted range interactions and mean--field models are hierarchical structures inspired by grammar \cite{DeGuili_2019a,DeGuili_2019b}, but these also involve interactions that reach explicitly over long distances.  It seems worth noting that in the absence of explicit positional encoding, the transformer architecture at the heart of LLMs is like a mean--field theory since it allows interactions among tokens independent of their separation so long as they fit into the allowed context~\cite{vaswani2017attention}.  

In summary, we have examined the predictability and information content of English text corpora through the lens of large language models.   Our analyses extend up to $10^{4}$ characters, longer than what has been accessible via human-based~\cite{Shannon_1951,Cover+King_1978} or smaller model-based methods~\cite{Brown1982estimate,noteL,Takahira2016entropy}.  We see consistent trends across four quite different but well trained, performant models. On the other hand we see significant differences across genres in the dependence of code length $L(N)$ and conditional entropy $s(N)$ on the context length $N$, and it is not clear whether this reflects inherent features of the real text or each genre's relationship to the models' training sets.  Nevertheless, a consistent trend is a continuous decrease in conditional entropy and code length over the full range of sequence lengths investigated. This continued decrease predicts the presence of correlations that span the entire length of the text, and for sufficiently large corpora  we can see this directly via measurements of mutual information. In addition we see a possibly diverging accumulation of short code length or low entropy tokens at large $N$. Although discovered through the use of LLMs, many of these results necessarily reflect the structure of language itself. Ongoing work on large context models, and assembling suitable companion corpora, suggest the possibility of pushing these analyses even further~\cite{gao2025train}.   It would be attractive to have toy models, in the spirit of statistical mechanics, that could explain these behaviors, even in outline.  

\begin{acknowledgments}
We thank K Krishnamurthy, G Reddy, DJ Schwab, and W Zhong for helpful discussions.  WB especially thanks MP Brenner, O Kimchi, FC Pereira, and the late N Tishby for conversations long ago that helped shape these questions.  This work was supported in part by the National Science Foundation through a Graduate Research Fellowship to LMS (Grant No. DGE-2039656) and by the Simons Foundation.
CS acknowledges support from the Center for the Physics of Biological Function and the Princeton Center for Theoretical Science. This work used the ACES cluster at Texas A{\&}M University through allocation CIS250478 from the Advanced Cyberinfrastructure Coordination Ecosystem: Services \& Support (ACCESS) program, which is supported by U.S. National Science Foundation grants \#2138259, \#2138286, \#2138307, \#2137603, and \#2138296, which was awarded as a part of the NSF Artificial and Natural Intelligence Institute. This work also used the Princeton Language and Intelligence computational resources.
\end{acknowledgments}

\appendix

\section{Corpora}
\label{app:corpora}

Scraping the internet generates a wide diversity of material, including segments which are not English text.  We restrict our analysis to strings that contain characters from a standard set chosen as 92 most common characters in the C4 internet corpus, plus the return carriage character $\setminus$r, which commonly appears in poetry. The resulting 93 characters are: 
\begin{quote}
$\setminus$r$\setminus$n\ !"\#\$\%\&'()*$+$,$-$./:;?–—‘’“”…[]\_

0123456789

ABCDEFGHIJKLMNOPQRSTUVWXYZ

abcdefghijklmnopqrstuvwxyz
\end{quote}

In constructing Fig.~\ref{fig01}, we sample strings randomly from the C4 data set. For each string, we first check that it contains only the characters from the set above.  
To randomize the starting point, the first $B$ characters of the string are deleted, where $B$ is randomly chosen between 0 and 100 for each string.  Next, the cropped string is  tokenized and the first $N_T=4096$ tokens are used for code length calculations; strings  shorter than 4096 tokens were omitted. Each curve represents an average over 30,000  strings. We only train our DCLM~1.7B model on strings of length $N_T=2048$, so we used $N_T=2048$ in the testing procedure for this model. 

For Figure \ref{fig02} an identical methodology was used for the C4 dataset. For the English Wikipedia and Gutenberg Poetry Corpus, one additional step was applied: sufficiently long  sequences of tokens were divided to accommodate multiple disjoint strings of length $N_T$. For English Wikipedia and C4, we use $N_T=4096$, and for the Gutenberg Poetry Corpus (GPC), we used $N_T=2048$. The C4 curve is an average over 61,280 independent strings, the English Wikipedia curve is an average over 83,360 independent strings, and the GPC curve is an average over 6,220 independent strings. Figure \ref{fig03} is generated from the same data as the C4 curve in Fig.~\ref{fig02}. Figure \ref{fig04} uses 78,960 independent strings from the C4 dataset.

For Figure~\ref{Id}, the empirical joint distribution $Q_d(c,c')$ is computed as follows. First, strings are broken into segments of length $N_C$ and strings shorter than $N_C$ are omitted. Within each segment, all pairs of the form $(c_{n},c_{n+d})$ for $n=1,\dots,N_C-d$ are used to build to the empirical joint distribution. For English Wikipedia, we use $N_C=5,000$ and for C4, we use $N_C=37,000$. If a string is long enough to accommodate multiple segments of length $N_C$, in the C4 dataset we only use the first, while in the Wikipedia dataset we use all of them. 

For Figure~\ref{fig:fig2Tokens} in Appendix \ref{app:data}, below,  we split each string into segments of no more than 15,000 characters.  We then sort the all the segments by length and retain the longest 2,000 segments.  Of these segments, the first $B$ characters are eliminated, where $0< B < 100$ is chosen randomly and independently for each segment, as before. These segments are tokenized out to a maximum length of 2,000 tokens. For the C4 dataset, we perform this pipeline on a 3--file subset of the 1024--file dataset. For English Wikipedia, we perform this analysis on a $100,000$--article subset of the full corpus. For the remaining datasets, the full corpus is used. 

\section{More about models}
\label{app:models}

We use and compare multiple models in the hope of uncovering properties of natural language rather than features of the models themselves. 

{\bf Llama 3.2 1B}
was released in 2024 by Meta. This model was pre-trained on up to 9 trillion multilingual tokens with a knowledge cutoff of December 2023 \cite{grattafiori2024llama3herdmodels}. The model has 1.23 billion parameters and uses Grouped-Query Attention (GQA) with a context length of 128,000 tokens. Additionally, logits from Llama 3.1 8B and 70B were incorporated into the pre-training stage: these logits were used as token-level targets. This model was post-trained and tuned into a final chat model via Supervised Fine-Tuning (SFT), Rejection Sampling (RS), and Direct Preference Optimization (DPO).

{\bf OLMo 2 1B}
was released in 2025 by the Allen Institute for AI. The model was pre-trained on 4 trillion tokens with a knowledge cutoff in December 2023 \cite{olmo20242olmo2furious}. Pre-training used the OLMo-mix-1124 dataset, followed by mid-training on Dolmino-mix-1124. The architecture comprises 1 billion parameters across 16 layers, with a model dimension of 2048 and 16 attention heads. It supports a context length of 4,096 tokens. For this work, we use the base model after pre- and mid-training, excluding the instruction-finetuned version that underwent additional post-training steps: supervised fine-tuning (SFT) on an OLMo-specific variant of the Tülu 3 dataset, further DPO training on olmo-2-0425-1b-preference-mix, and final reinforcement learning with verifiable rewards.

{\bf Qwen3~8B} was released in 2025 by Alibaba. This model was pre-trained on 36 trillion multilingual tokens with an unknown knowledge cutoff date, but probably some time in 2024 \cite{qwen3technicalreport}. The model has 8.2 billion parameters consisting of 36 layers and GQA with 32 query heads and 8 key-value heads.
It has a native context length of 32,768 tokens.
This model was post-trained using Strong-to-Weak Distillation from either the Qwen3 32B model or the Qwen3 235B-A22B model. The distillation is comprised of a first phase of Off-Policy Distillation and a second phase of On-Policy Distillation.

{\bf DCLM~1.7B} is a model that we trained ourselves using the recipe for a 1B model on DCLM's \cite{li2024datacomplm} Github repository \cite{Mlfoundations}. For this study, we perform only a basic pre-training procedure on 28B tokens of this data and do not perform any post-training or instruction finetuning. We use ``\texttt{global-shard\_03\_of\_10/local-shard\_1\_of\_10}" from DCLM-baseline as our data, which has a knowledge cutoff of 2022. We tokenize and shuffle the data using DCLM's Rust code and the EleutherAI gpt-neox-20b tokenizer \cite{black2022gptneox}, resulting in sequences of length 2048 tokens. These training data are a filtered subset of the C4 data. Our model is an adaptation of the vanilla Transformer \cite{vaswani2017attention} model from NanoGPT \cite{nanogpt}. It has a context length of 2048 tokens, 24 layers, 16 attention heads with QK-norm \cite{henry2020querykeynormalizationtransformers}, and model dimension 2048. We use absolute positional embeddings, imposing a maximum context length of 2048 tokens. We trained using a seed of 1337 for both the model weight random initializations and the dataloader shuffling. We used a batch size of 6 with 48 gradient accumulation steps, training for 47471 steps on 8 H100 GPUs for approximately 30 hours on each GPU. We used AdamW \cite{loshchilov2017decoupled} as our optimizer with default hyperparameters, and a standard cross-entropy loss function with an additional z-loss hyperparameter set to $10^{-4}$. Our weight decay was set to 0.033 and we used a cosine-decay with linear warm-up learning rate schedule, with a minimum learning rate of $3\times 10^{-4}$, maximum learning rate of $3\times 10^{-3}$, and $5000$ warm-up steps.

\section{Data summary}
\label{app:data}

Different models use different tokenizers, and if we are not careful this would mean that each model uses an idiosyncratic unit of length, even differing across corpora.  
Table~\ref{tab:KtoN} contains the average number of characters per token for each dataset and model. Our analyses of code length and conditional entropy have been repeated across all 36 combinations of models and corpora described above.  Results are summarized in
Fig.~\ref{fig:fig2Tokens}.

\begin{table}[b]
\centering
\resizebox{\linewidth}{!}{
\begin{tabular}{|l|c|c|c|c|}
\hline
 \textbf{Dataset}           &  \textbf{OLMo~2 1B} &  \textbf{Llama~3.2 1B} & \textbf{DCLM~1.7B} &  \textbf{Qwen3~8B} \\
\hline
Poetry Foundation Poems     & 4.16 & 4.16&3.75 & 4.16 \\
Gutenberg Poetry Corpus     & 3.93 & 3.92&3.59 & 3.92 \\
Narrative Content           & 4.03 & 4.03&3.96 & 4.00\\ 
Narrative Analyses          & 4.70 & 4.70&4.64 & 4.70 \\
Narrative Summaries         & 4.48 & 4.48&4.41 & 4.48 \\
C4 Internet Corpus          & 4.79 & 4.79&4.65 & 4.72 \\
BBC News Articles           & 4.73 & 4.75&4.55 & 4.67 \\
English Wikipedia           & 4.61 & 4.61&4.54 & 4.46 \\
Simple English Wikipedia    & 4.44 & 4.44&4.32 & 4.30 \\
\hline
\end{tabular}}
\caption{Average number of characters per token across all combinations of models and corpora.}
\label{tab:KtoN}
\end{table}

\setcounter{figure}{0}
\renewcommand{\thefigure}{A\arabic{figure}}
\begin{figure*}
    \centering
    \includegraphics[width=0.9\textwidth]{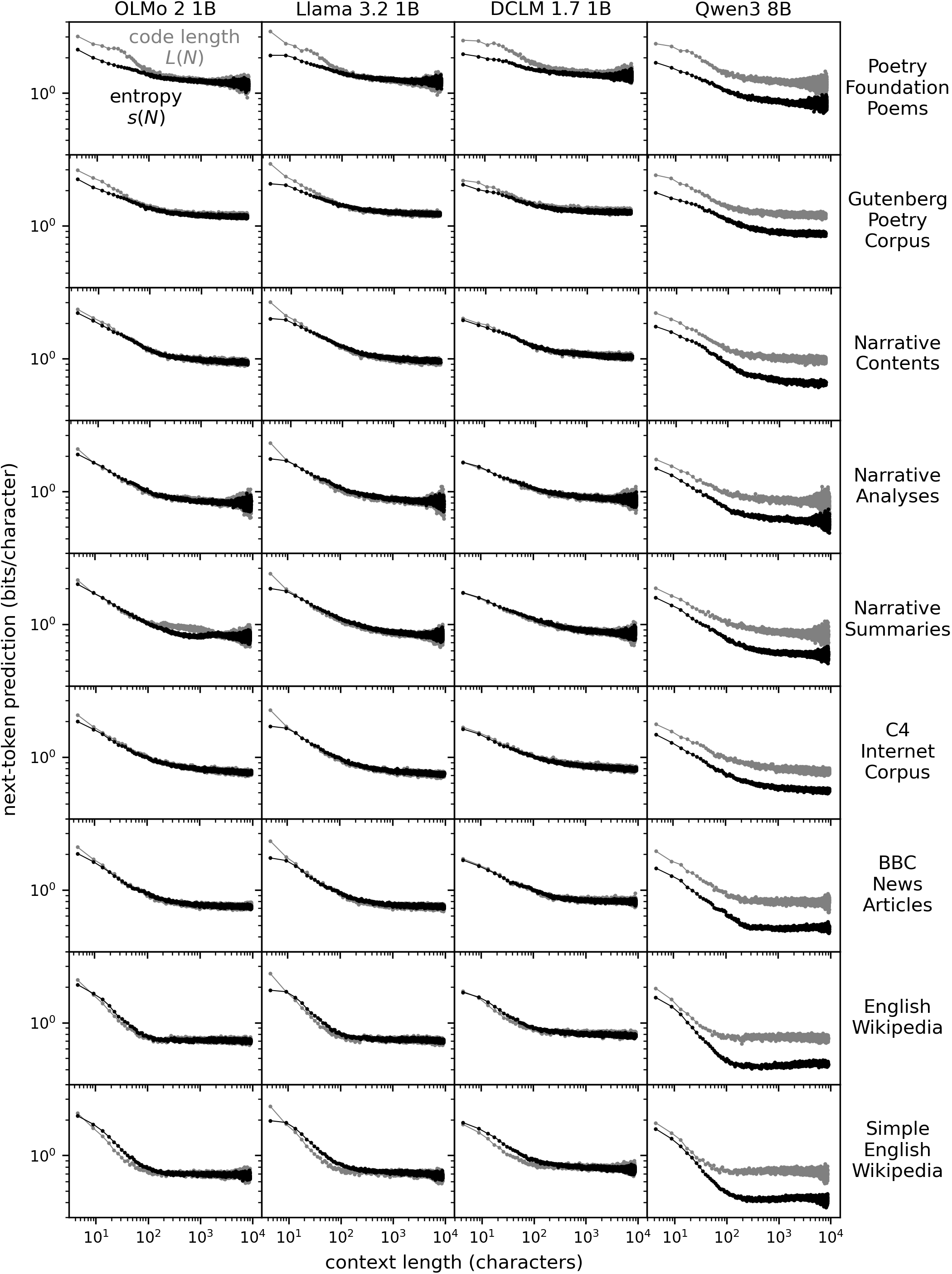}
    \caption{
    {\bf Conditional entropy and code length across models and genres.} 
    The mean conditional entropy $s(N)$ (black) and code-length $L(N)$ (gray) for all combinations of corpora and models that we explored. Different models use different tokenization schemes but we use characters as the unit of length, converting through the mean $N/K$ from Table \ref{tab:KtoN}.
    \label{fig:fig2Tokens}}
\end{figure*}

\bibliography{LLM}
\end{document}